\documentclass[letterpaper, 10 pt, conference]{ieeeconf}  

\IEEEoverridecommandlockouts                              

\usepackage{tikz}
\usepackage{pgfplots}
\usepackage{graphicx} 
\usepackage{setspace}
\usepackage{xcolor}
\usepackage{amsmath}
\usepackage{amsfonts}
\usepackage{bbm}
\usepackage{mathtools} 
\RequirePackage{algorithm}
\RequirePackage{algpseudocode} 
\usepackage{caption}
\usepackage{subcaption}
\usepackage{comment}
\usepackage{listings}
\usepackage{dsfont}
\usepackage{amsfonts}
\usepackage{optidef}
\usepackage{hyperref}
\usepackage{url}
\newcommand{\norm}[1]{\left\lVert#1\right\rVert}
\newcommand{\R}[1]{\mathrm{#1}}		
\newcommand{\C}[1]{\emph{#1}}

\newcommand{\fdl}[1]{{\textbf{\color{teal} FDL: #1}}}

\definecolor{myorange}{RGB}{205, 102, 57}
\pgfplotsset{compat=1.18} 

\usepackage{csquotes}

\title{\LARGE \bf
Emergent Cooperative Strategies for Multi-Agent Shepherding via Reinforcement Learning
}

\author{Italo Napolitano\textsuperscript{1}, Andrea Lama\textsuperscript{1}, Francesco De Lellis\textsuperscript{2}, Mario di Bernardo\textsuperscript{1, 2, *}
\thanks{This work was developed with the economic support of MIUR (Italian Ministry of University and Research) performing the activities of the project PRIN 2022 “Machine-learning based control of complex multi-agent systems for search and rescue operations in natural disasters (MENTOR).}
\thanks{\textsuperscript{1}Scuola Superiore Meridionale, Naples, Italy}%
\thanks{\textsuperscript{2}Department of Electrical Engineering and
Information Technology, University of Naples Federico II, Naples, Italy}%
\thanks{\textsuperscript{*}Corresponding author \href{mailto:mario.diberanardo@unina.it}{mario.dibernardo@unina.it}}%
\thanks{The authors wish to thank Mr Stefano Covone, MSc student in Automation and Robotics Engineering at the University of Naples Federico II, for providing Figure \ref{fig:subtasks} of the manuscript.}%
}

\begin{document}

\maketitle
\thispagestyle{empty}
\pagestyle{empty}

\begin{abstract}
We present a decentralized reinforcement learning (RL) approach to address the multi-agent shepherding control problem, departing from the conventional assumption of cohesive target groups. 
Our two-layer control architecture consists of a low-level controller that guides each herder to contain a specific target within a goal region, while a high-level layer dynamically selects from multiple targets the one an herder should aim at corralling and containing. Cooperation emerges naturally, as herders autonomously choose distinct targets to expedite task completion. We further extend this approach to large-scale systems, where each herder applies a shared policy, trained with few agents, while managing a fixed subset of agents. 
\end{abstract}

\section{Introduction}
The shepherding control problem exemplifies how collective behavior in complex systems can be leveraged to accomplish targeted tasks. It generally involves two agent groups: the \C{herders}, who coordinate to steer the overall dynamics of the \C{targets} towards a desired configuration. In control theory, shepherding is often viewed as an indirect control problem \cite{licitra2018}, as the objective of influencing the targets' dynamics is achieved by optimally managing the herders. This problem has wide-ranging applications across robotics \cite{sebastian2022}, crowd dynamics \cite{albi2016}, and more.

Given the complexity of the shepherding problem, obtaining an analytical solution is highly challenging and often necessitates significant simplifying assumptions \cite{pierson2018}. To approximate optimal strategies, several studies have introduced learning-based approaches \cite{long2020}, frequently relying on the assumption of cohesive target behavior, which simplifies the problem considerably \cite{go2016, hasan2022}. Our approach instead seeks to (i) relax the assumption of target cohesion, (ii) minimize reliance on heuristic assumptions to drive the herder behavior \cite{strombom2014}, and (iii) formulate the problem in an optimization setting that can be solved using learning-based methods.


As common in the literature \cite{delellis2021a, auletta2022a}, we decompose the shepherding task into two sub-tasks: a \C{driving}\footnote{In \cite{strombom2014}, \C{driving} refers to guiding the center of mass of a group of cohesive targets. In contrast, here we refer to guiding each target individually.} strategy, where a herder drives a target to a desired location; and a \C{target selection} strategy, where each herder selects a target to pursue.
We propose a decentralized, learning-based policy for each sub-task and achieve scalability by limiting each herder’s interactions to a fixed number of target agents. Our results show that herders autonomously learn to cooperate, efficiently completing the task by selecting distinct targets.

\subsection{State of the art} 
Given the complexity of the shepherding problem, a fully control-based solution for a general scenario remains elusive without specific assumptions \cite{sebastian2022, licitra2018}. Both heuristic and control-based methods can be suboptimal or encounter challenges in complex, time-varying environments with heterogeneous agents \cite{vanhavermaet2024}. Consequently, learning-based approaches, particularly Reinforcement Learning (RL), are increasingly being explored to approximate optimal control strategies \cite{sutton2020}.
A crucial distinction exists between single and multi-herder scenarios. Single-herder problems involve one learning agent, while multi-herder systems require learning in shared environments, typically tackled using Multi-Agent Reinforcement Learning (MARL) \cite{gupta2017}.

In RL-based single-herder cases, most approaches simplify the problem by assuming cohesive targets, such as flocking agents. This assumption significantly simplifies the problem by, for instance, reducing the targets' dynamics to that of their center of mass \cite{go2016}.  
Additionally, several solutions incorporate heuristics, e.g. where herders learn to switch between predefined behaviors \cite{hussein2022}.

Only a few studies have employed learning-based methods for multiple herders. For instance, \cite{patil2023} models human-like behaviors using Dynamical Perceptual-Motor Primitives (DPMP) to represent herders, with a high-level controller for target selection trained via Proximal Policy Optimization (PPO). In \cite{delellis2021a}, control-tutored reinforcement learning (CTRL) is proposed to reduce training time for tabular RL agents. Though this work assumes non-cohesive targets, it focuses on training a single herder for one target, then scales to multiple agents by employing a heuristic target-selection mechanism. 
Similar to the single-herder case, current multi-agent learning-based shepherding solutions typically assume cohesive forces among targets \cite{hasan2022, wang2024}. In this work, we propose a learning-based solution for multi-agent shepherding that eliminates the cohesion assumption and minimizes reliance on heuristics, addressing a key gap in the literature.

\section{Problem statement}
We consider a dynamical system involving two interacting populations in a two-dimensional space. 
The first group, referred to as \textit{herders}, consists of $n$ controlled agents; the second group, called \textit{targets}, includes $m$ passive agents.\\
The objective is to design a strategy for the herders to steer and contain the targets within a \emph{goal region} in the plane. 

Without loss of generality, we define a square domain $\mathcal{D} = \left[-L, L\right]^2 \subset \mathbb{R}^2$ and a circular goal region $\Omega_{\text{G}} \subset \mathcal{D}$ with radius $\rho_{\text{G}} < L$.
We denote the position of the $i$-th target as $\mathbf{T}_i \in \mathcal{D}$ and that of the $j$-th herder as $\mathbf{H}_j \in \mathcal{D}$. 
For brevity, we represent the stacked position coordinates of all targets and herders as $\mathbf{T}$ and $\mathbf{H}$, respectively.

We model the herders' behavior using first-order differential equations and the targets' behavior with second-order differential equations, in the spirit of \cite{albi2016}. 
However, unlike typical approaches in the literature, we relax the cohesiveness assumption and define the dynamics of the $i$-th target as a second-order Langevin equation of the form:
\begin{equation}
    \mathbf{\Ddot{T}}_i (t) = - \zeta \mathbf{\dot{T}}_i (t) + c\mathbf{i}^\R{HT}_i\left(\mathbf{H}(t), \mathbf{T}_i(t)\right) + \sigma \mathbf{N}_i (t),
    \label{eq:targets_dyn_TC}
\end{equation}
which includes a damping term $-\zeta \dot{\mathbf{T}}_i(t)$, where $\zeta > 0$ is the damping coefficient; a white Gaussian noise vector $\mathbf{N}_i \in \mathbb{R}^2$ with a diffusion coefficient $\sigma > 0$, and an interaction term $\mathbf{i}^{\text{HT}}_i\left(\mathbf{H}(t), \mathbf{T}_i(t)\right)$ describing the influence of nearby herders, with interaction strength $c > 0$. 
The latter is defined as
\begin{multline} 
\mathbf{i}^\R{HT}_i\left(\mathbf{H}, \mathbf{T}_i\right) = \\
= \frac{1}{2} \sum_{j=1}^m \left( 1 - \tanh \left( \frac{\beta (\norm{\mathbf{T}_i - \mathbf{H}_j} - \lambda)}{\lambda} \right)\right) \frac{\mathbf{T}_i - \mathbf{H}_j}{\norm{\mathbf{T}_i - \mathbf{H}_j}}, 
\label{eq:interaction_kernel} 
\end{multline}
which describes a repulsion force exerted on the $i$-th target by nearby herders. This smoothly decays to zero after a typical distance $\lambda > 0$ \cite{sebastian2022}, with the stiffness of the decay being controlled by the parameter $\beta > 0$. These forces are additive when multiple herders interact with a target. 

The $j$-th herder, on the other hand, is a single integrator of the exogenous control action $\mathbf{u}_j$, as follows
\begin{equation}
    \mathbf{\dot{H}}_j(t) = \mathbf{u}_j(t).
    \label{eq:herders_dyn_TC}
\end{equation}

For computational reasons and to accommodate the discrete-time nature of controllers and actuators, we formulate the shepherding control problem as the following discrete-time optimal control problem:
\begin{maxi!}|s|[3]
    {\pi}{J= \mathbb{E} \left[\sum_{k=0}^{N_\R{h}}\gamma^k r_k(\mathbf{T}, \mathbf{H}) \right] \label{eq:obj_fcn}}{\label{eq:opt_prob}}{}
    \addConstraint{\mathbf{H}_j(k+1)=\mathbf{H}_j(k) + \mathbf{u}_j(k) \Delta t,\, \forall j \in [1, n] \label{eq:herders_dyn}}
    \addConstraint{\mathbf{T}_i(k+1)=\mathbf{T}_i(k) + \mathbf{V}_i(k) \Delta t,\, \forall i \in [1, m] \label{eq:targets_dyn}}
    \addConstraint{\mathbf{V}_i(k+1)= f_\R{T}\left(\mathbf{H}(k), \mathbf{T}_i(k), \mathbf{V}_i(k) \right),\, \forall i \in [1, m] \label{eq:targets_vel}}
    \addConstraint{\mathbf{u}_j(k) \sim \pi \left(\cdot \mid \mathbf{T}(k), \mathbf{V}(k), \mathbf{H}(k) \right), \forall j \in [1, n] \label{eq:act_sample}}
    \addConstraint{\mathbf{H}(0) = \mathbf{H}_{\R{0}} \text{, } \mathbf{T}(0) = \mathbf{T}_{\R{0}} \text{, } \mathbf{V}_i(0) = \mathbf{0}_{2} ,\, \forall i \in [1, m] \label{eq:initial_conditions}}
    \addConstraint{\mathbf{T}_i, \mathbf{H}_j \in \mathcal{D} ,\, \forall j \in [1, n] ,\, \forall i \in [1, m] \label{eq:state_space_1}}
    \addConstraint{\mathbf{V}_i \in \left[-V_\R{max,T}, V_\R{max,T}\right]^2 ,\, \forall i \in [1, m] \label{eq:state_space_2}}
    \addConstraint{\mathbf{u}_j \in \left[-V_\R{max,H}, V_\R{max,H}\right]^2,\, \forall j \in [1, n] \label{eq:action_space}}
\end{maxi!}
where $\pi$ is the optimal control policy to be determined, $\gamma \in \left(0, 1\right)$ represents the discount rate, and $r_k$ is the reward function at time step $k$, which will be defined later. The parameter $N_{\text{h}}$ represents the time horizon. Moreover, we denote the $i$-th target's velocity as $\mathbf{V}_i$, limited by $V_{\text{max,T}}$ for each target, while the herders have a velocity limit of $V_{\text{max,H}} > V_{\text{max,T}}$ \cite{long2020}. We represent the stacked velocity of all target agents as $\mathbf{V}$.


Eqs \eqref{eq:herders_dyn}--\eqref{eq:targets_vel} represent the discretized model of the dynamical system presented in Eq. \eqref{eq:targets_dyn_TC}--\eqref{eq:herders_dyn_TC}, where
\begin{equation}
\begin{split}
&f_\R{T}\left(\mathbf{H}(k), \mathbf{T}_i(k), \mathbf{V}_i(k)\right) = \\
&= \mathbf{V}_i(k) + \left( - \zeta \mathbf{V}_i(k) + c \mathbf{i}^\R{HT}_i\left(\mathbf{H}(k), \mathbf{T}_i(k)\right) \right) \Delta t + \\
& \ \ \ + \sigma \mathbf{N}_i(k) \sqrt{\Delta t},
\label{eq:target_full_dyn} 
\end{split}
\end{equation}
and where the control action $\mathbf{u}_j(k)$ is sampled from the policy $\pi \left(a \mid \mathbf{T}(k), \mathbf{V}(k), \mathbf{H}(k) \right)$ in \eqref{eq:act_sample}, which represents the probability of taking action $a$ given the current state.

The initial conditions are specified in Eq. \eqref{eq:initial_conditions}, while Eqs. \eqref{eq:state_space_1}, \eqref{eq:state_space_2}, and \eqref{eq:action_space} constrain the states and actions to their respective domains.

\section{Control architecture}
Assuming the dynamics of the targets in Eqs. \eqref{eq:targets_dyn}--\eqref{eq:targets_vel} are unknown, we propose a DQL-based strategy to solve the optimization problem \eqref{eq:opt_prob}. For multiple herders, we adopt a policy-sharing protocol \cite{gupta2017}. 

\begin{figure}[tb]
    \centering
    {\includegraphics[width=0.45\textwidth]{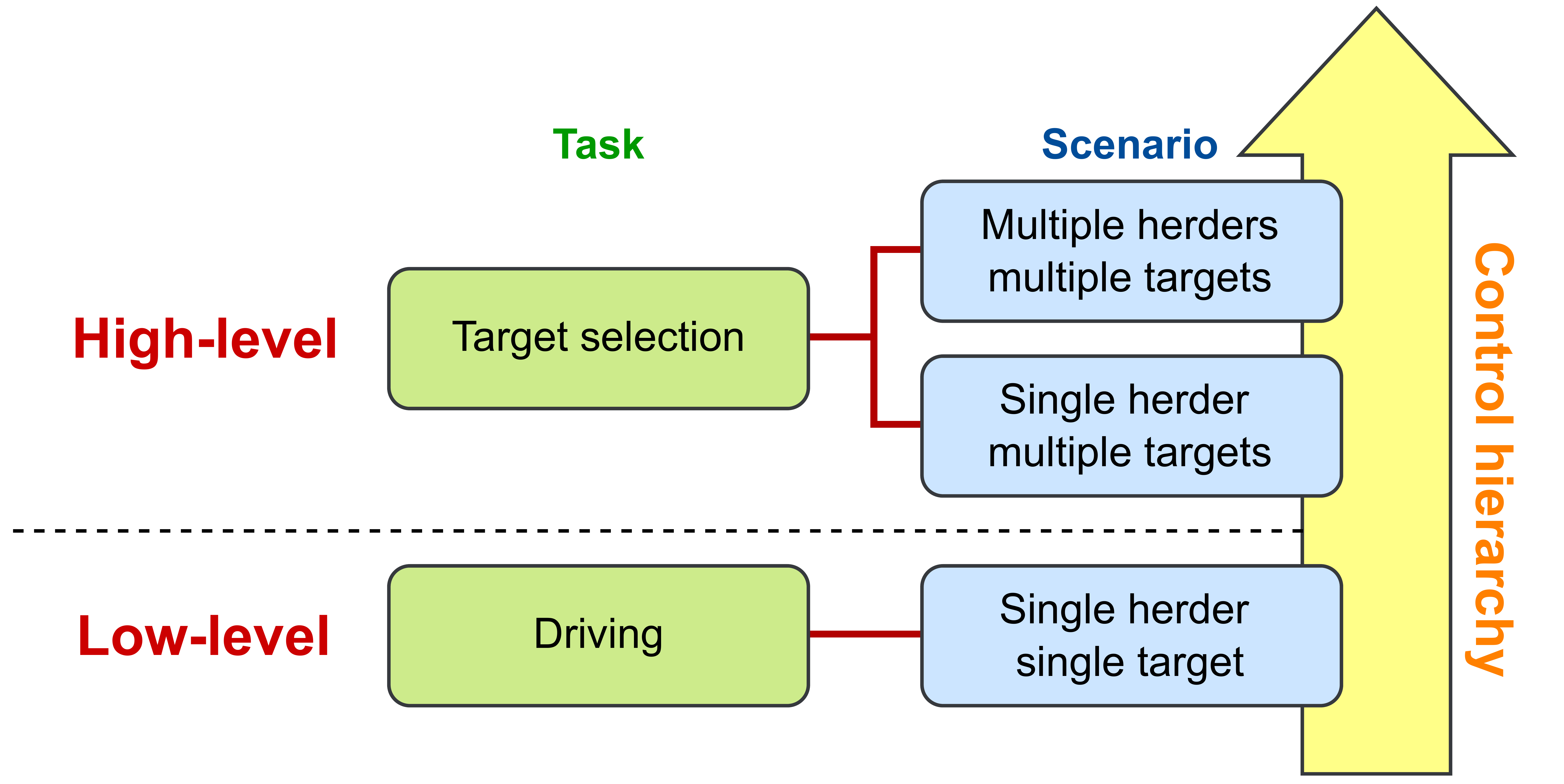}}
    \caption{Schematic representation of the decentralized two-layer control architecture. Each herder independently interrogates the high-level policy to select a target, which then informs the low-level control inputs needed to guide that target towards the goal region.}
    \label{fig:subtasks}
\end{figure}

Given the lack of cohesion among targets, herders must interact with each target individually. To address this, we decompose the problem into two sub-tasks (see \cite{delellis2021a, auletta2022a}): {\it driving} and {\it target selection}. This decomposition leads to a two-layer control architecture, shown in Figure \ref{fig:subtasks}.

The Deep Q-Network (DQN) training process encompasses three scenarios:
(i) single herder with single target,
(ii) single herder with multiple targets,
(iii) multiple herders with multiple targets.
In the first scenario, we train a low-level control for the driving sub-task. For the multiple-target scenarios, we focus on training the high-level control for target selection as the herders use the low-level control policy learned in the first scenario to steer the selected target. 

\section{Driving sub-task}
Consider the scenario with one herder and one target ($n=1, m=1$). In this case, the herder must learn to guide and contain the target from its initial position to the designated goal region, defining the sub-task we refer to as {\em driving}.

\subsection{Training} \label{sec:single_herder_single_target}
We simulate our model fixing nominal parameters as in Table \ref{tab:model_param}. 
We allow each episode to run for up to $N_\R{h} = 2000$ steps, terminating early if all targets remain within the goal region for $N_\R{t} = 200$ consecutive steps. The settling time, $\tau^{\star}$, is defined as the time at which all targets enter and remain within the goal region until the end of the episode. It is formally expressed as follows:
\begin{equation}
    \tau^{\star} = \inf \{ \tau \geq 0  \; \text{s.t.} \; \norm{\mathbf{T}_i(k)}_2 \leq \rho_\R{G}, \; \forall i, \forall k \in \left[ \tau, N_\R{f} \right] \},
    \label{eq:settling_time_def}
\end{equation}
where $N_\R{f} = \min \left( \tau + N_\R{t}, N_\R{h} \right)$.
Thus, we define an episode to be successful if a finite settling time exists.

To solve the driving sub-task using RL, we train a DQN that takes as inputs the herder's position, as well as the target's position and velocity. The outputs are the $x$ and $y$ components of the herder's discretized velocity vector. Since DQN supports a continuous state space but requires a discrete action space \cite{mnih2015}, we discretize each velocity component into five bins. The DQN consists of 6 input neurons, two hidden layers with 128 and 64 neurons respectively, both with ReLU activation, and 25 output neurons with linear activation.

We shape the reward function at time step $k$ as follows
\begin{multline}
    r_k = -k_1 \norm{\mathbf{T}(k)} - k_2 \norm{\mathbf{T}(k) - \mathbf{H}(k)} + \\
    + k_3 \cdot \mathds{1}_\R{\{\norm{\cdot} \leq \rho_\R{G}\}}\left(\mathbf{T}(k)\right) - k_4 \cdot \mathds{1}_\R{\{\norm{\cdot} \leq \rho_\R{G}\}} \left(\mathbf{H}(k)\right)
    \label{eq:reward_1h1t}
\end{multline}
where $k_i > 0 \; \forall i=\{1,\dots,4\}$, and $\mathds{1}_{\{\R{A}\}}(x)$ is the indicator function, equal to $1$ if $x \in \R{A}$ and $0$ otherwise.
The first term penalizes the distance between the herder and the target, encouraging the herder to approach the target. The second term penalizes the target's distance from the origin, motivating the herder to steer the target towards the goal region at the center of the plane. The third and fourth terms provide sparse rewards to encode the final goal: a positive reward if the target enters the goal region and a penalty if the herder enters it, useful during containment.

To train the DQN agent, we run a maximum of $E = 2000$ episodes, using the hyperparameters reported in Table \ref{tab:hyperparameters}. 
To prevent the agent from learning to steer targets located only in specific regions of the plane, we randomize the initial targets' positions. These positions are generated uniformly across all four quadrants of the plane, ensuring comprehensive exploration of the state space. Moreover, herder's initial position is randomly sampled uniformly across the plane, independently of the target's position.

During training, we use an early stopping criterion as defined in \cite{delellis2022}. 
Specifically, training terminates when the agent successfully completes 10 consecutive episodes in each of the four quadrants, totaling 40 consecutive successful episodes overall (see Sec. \ref{sec:single_herder_single_target_val}). 
We denote the episode number at which this condition is met as $E_{\text{t,40}}$. 

\begin{figure}[tb]
    \centering
        \begin{tikzpicture}[remember picture]
        \node[anchor=south west,inner sep=0] (image) at (0,0) {\includegraphics[width=0.4\textwidth, height=0.22\textwidth, trim={4.5cm 3cm 0cm 0cm}, clip]{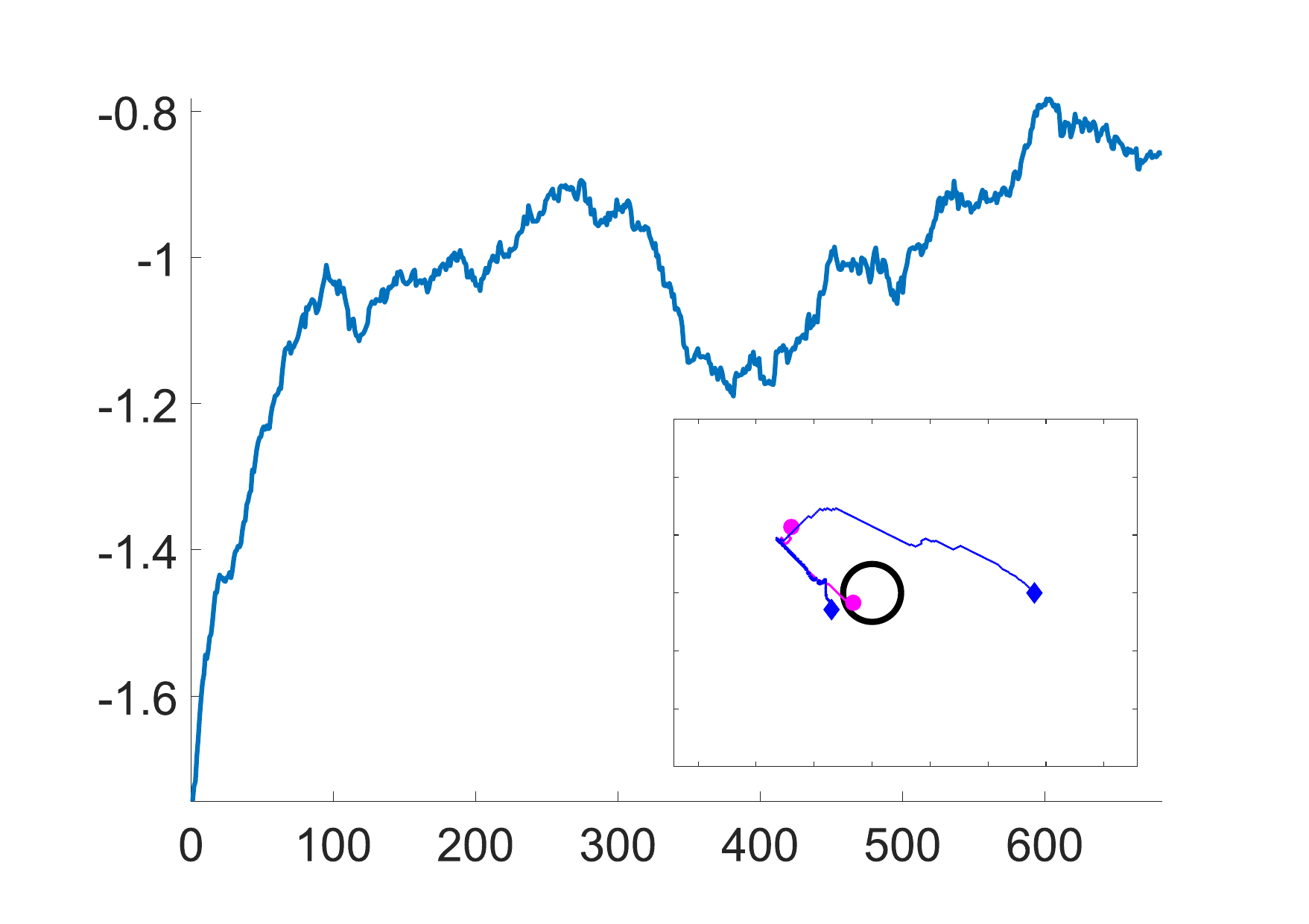}};
        \begin{scope}[x={(image.south east)},y={(image.north west)}]
            \draw[-] (0,0) -- (0.87,0) ;
            \draw[-] (0,0) -- (0,0.87) node[above] {\quad x$10^5$};

            \foreach \x/\xlabel in {0/0, 0.145/100, 0.290/200, 0.435/300, 0.580/400, 0.725/500, 0.87/600}
                \draw (\x,0) -- (\x,-0.02) node[below] {$\xlabel$};
            \foreach \y/\ylabel in {0/-1.8, 0.175/-1.6, 0.350/-1.4, 0.525/-1.2, 0.700/-1, 0.87/-0.8}
                \draw (0,\y) -- (-0.02,\y) node[left] {$\ylabel$};

            \node at (0.45,-0.2) {Episodes};
            \node[rotate=90]  at (-0.18,0.45) {Cumulative reward};
        \end{scope}
            \node[] at (6.18,3.75) {(a)};
            \draw[->] (6.18,3.5) -- ++(0, -0.25) node[below] {};        
            \node[] at (3.4,0.4) {(a)};
    \end{tikzpicture}
    \caption{Cumulative rewards of the training of the $n=1, m=1$ scenario. The curve is smoothed with a moving average of $100$ steps (taken on the left). The inset (a) shows a validation example of the policy saved at the episode indicated by the red arrow. The herder (blue diamond) corrals the target (magenta dot) and steers it inside the goal region (black circle). The trajectories of the herder and target are shown by blue and magenta small dots, respectively, while their final positions are indicated by the blue diamond and magenta dot.}
    \label{fig:1h1t_training_val}
\end{figure}
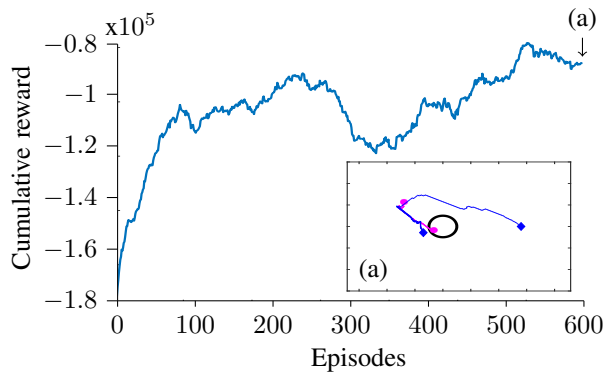

The results of the training procedure, obtained from numerical simulations, are summarized by the learning curve depicted in Figure \ref{fig:1h1t_training_val}. Notably, the agent meets the early stopping condition at episode $E_{\text{t,40}} = 683$.

\subsection{Validation}\label{sec:single_herder_single_target_val}
We keep $N_\R{h} = 2000$ and $N_\R{t} = 200$ and validate the learned policy using 1000 random initial conditions, uniformly distributed within a circle of radius $L/2$. The policy achieves a 100\% success rate, with an average settling time of $\tau^\star = 682 \pm 320$ steps (see Table \ref{tab:validation}). 
The inset in Figure \ref{fig:1h1t_training_val} shows an example of typical trajectories learned by the model, where the herder approaches and corrals the target to the goal region.


\section{Target selection sub-task}
Consider the scenario of a single herder and multiple targets. After learning to steer a target, the herder must also learn to select which target to pursue from multiple candidates.
Due to the fixed structure of the DQN, we assume that each herder is limited to selecting from a maximum of $\hat{m} > 1$ targets, regardless of the total number present, which presents a scalability challenge addressed in Section \ref{sec:scalability}.
This setup results in $2(1 + \hat{m})$ neurons in the input layer (representing the positions of the herder and of nearby targets) and $\hat{m}$ neurons in the output layer (returning the Q-value of each target selection choice). The selected target index then defines the inputs for the low-level control, guiding the herder to push the chosen target.



\subsection{Training}
We consider the scenario of a single herder and multiple targets, with $\hat{m} = m = 5$. The model parameters are set as in Table \ref{tab:model_param}. The DQN takes the positions of the agents as inputs, and its action space consists of selecting target indices. The network includes two hidden layers with 512 and 256 neurons, respectively, using ReLU activation functions for the hidden layers and linear activation function in the output layer. Training hyperparameters are reported in Table \ref{tab:hyperparameters}, with $N_\R{h} = 10000$, $N_\R{t} = 200$, and $E = 20000$.

We define and shape a reward function as follows:
\begin{equation}
    r_k = - k_5 \sum_{i=1}^m \text{ReLU} \left(\norm{\mathbf{T}_i (k)}  - \rho_\R{G} \right)
    \label{eq:reward_nhmt}
\end{equation}

This function penalizes the cumulative distance of all targets from the origin when they are outside the goal region, implicitly encouraging minimization of the settling time. During training, the target selection decision is kept fixed for a window of $n_{\text{w}}$ time steps, meaning high-level control is applied every $n_{\text{w}}$ steps. However, during validation, this constraint is removed, allowing the herder to switch targets freely throughout the episode.

\begin{figure}[tb]
    \centering
        \begin{tikzpicture}[remember picture]
        \node[anchor=south west,inner sep=0] (image) at (0,0) {\includegraphics[width=0.4\textwidth, height=0.16\textwidth, trim={4.5cm 3cm 0cm 0cm}, clip]{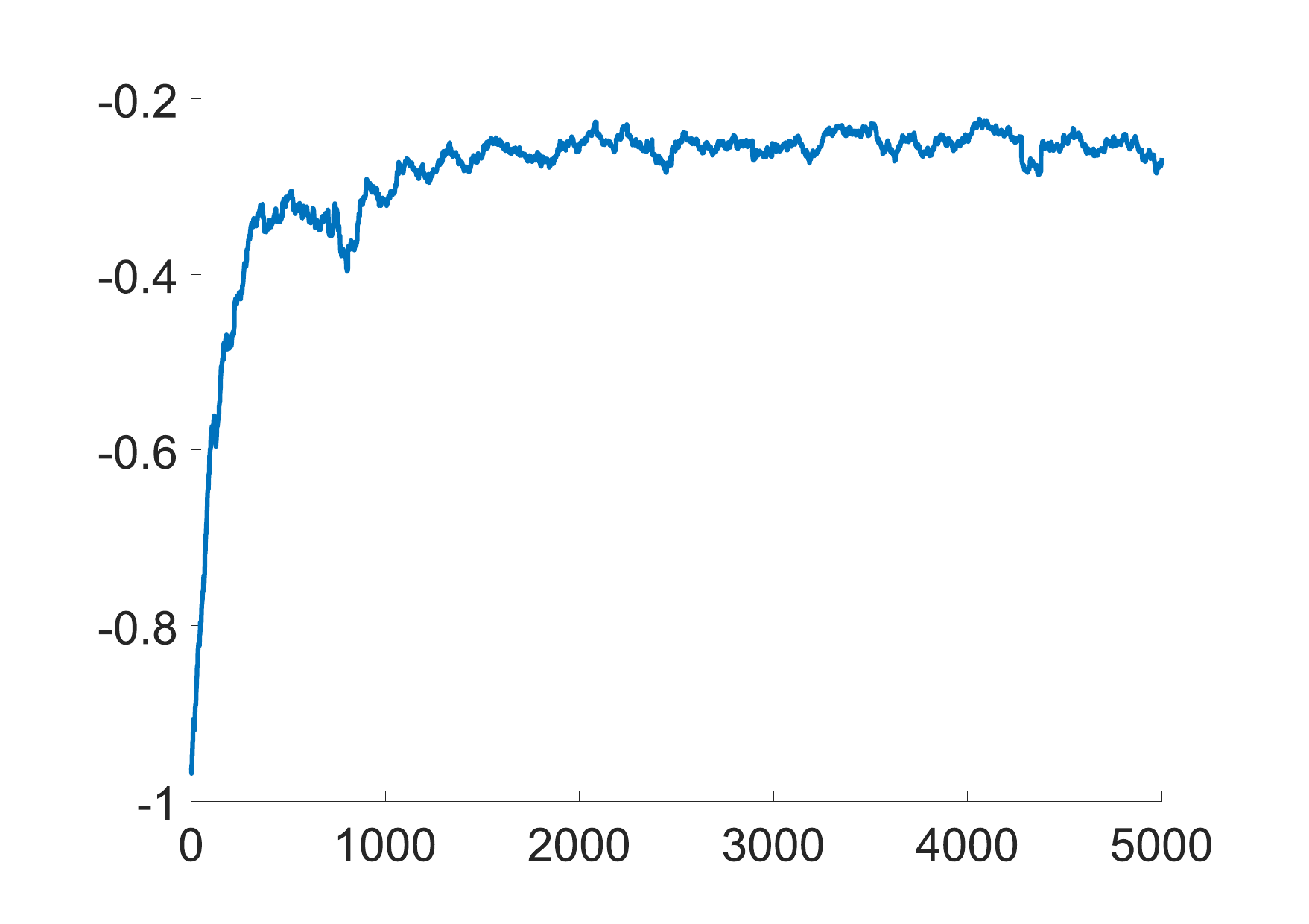}};
        \begin{scope}[x={(image.south east)},y={(image.north west)}]
            \draw[-] (0,0) -- (0.87,0) ;
            \draw[-] (0,0) -- (0,0.87) node[above] {\quad x$10^4$};

            \foreach \x/\xlabel in {0/0, 0.174/1000, 0.348/2000, 0.522/3000, 0.694/4000, 0.87/5000}
                \draw (\x,0) -- (\x,-0.02) node[below] {$\xlabel$};
            \foreach \y/\ylabel in {0/-1, 0.2175/-0.8, 0.435/-0.6, 0.6525/-0.4, 0.87/-0.2}
                \draw (0,\y) -- (-0.02,\y) node[left] {$\ylabel$};

            \node at (0.45,-0.24) {Episodes};
            \node[rotate=90]  at (-0.18,0.45) {Cumulative reward};
        \end{scope}
    \end{tikzpicture}
    \caption{Cumulative reward curve of the training of the $n=1, \, m=5$ scenario. For readability purposes, the curve is smoothed with a moving average of $100$ steps (taken on the left) and we show only the first $5000$ episodes of the training.}
    \label{fig:1hmt_training}
\end{figure}
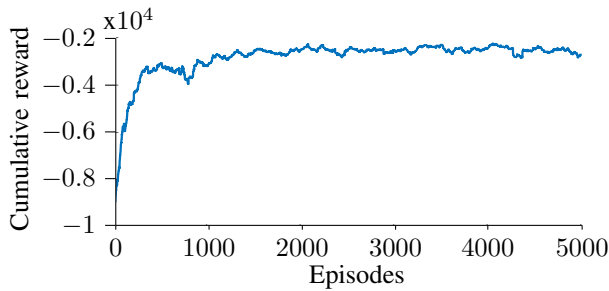

Figure \ref{fig:1hmt_training} shows the cumulative reward that gradually converges to a steady value during training. 
Notably, by leveraging the low-level policy presented in Sec. \ref{sec:single_herder_single_target} with $n_\R{w} = 100$, the herders can achieve successful episodes even in the early stages of training. However, this does not imply the discovery of an efficient policy or its effectiveness when $n_\R{w} = 1$. Therefore, we do not implement early stopping and training continues up to $E$ episodes, to encourage exploration of state space and improve policy performance.

\subsection{Validation}
We validate the trained policy over 1000 episodes, with random initial positions for all agents uniformly distributed within a circle of radius $L/2$. The target selection window is set to $n_{\text{w}} = 1$, allowing herders to choose a new target at each time step. As in training, we fix the number of herders and targets to $n=1$ and $m=5$, with $N_{\text{h}} = 10000$ and $N_{\text{t}} = 200$. Results, shown in Table \ref{tab:validation}, indicate a success rate of 92.8\%, with settling time of $\tau^* = 10088 \pm 3961$ steps.

\section{Multiple herders, multiple targets scenario}
Now consider the full setting of the proposed shepherding problem, involving multiple herders and multiple targets. 


We use Deep Q-Learning (DQL) in a multi-agent setting with parameter sharing to reduce the computational cost of training separate networks for each of the $n$ herders, assuming homogeneous dynamics \cite{gupta2017}. During training, all herders add transition vectors (comprising their observations, actions, and the global reward) to a shared replay buffer.

We assume each herder can sense a fixed number of herders $\hat{n} \geq 1$ and targets $\hat{m} > 1$, regardless of the total number of herders $n$ and targets $m$ in the environment. 
We adopt the model parameters from Table \ref{tab:model_param}, with a fixed number of agents $\hat{n} = n = 2$ and $\hat{m} = m = 5$. The hidden layer structure and hyperparameters are set as in the previous scenario (see Table \ref{tab:hyperparameters}), with $N_{\text{h}} = 10000$, $N_{\text{t}} = 200$, and $E = 10000$. The input layer consists of $2(\hat{n} + \hat{m})$ neurons, while the output layer has $\hat{m}$ neurons. 

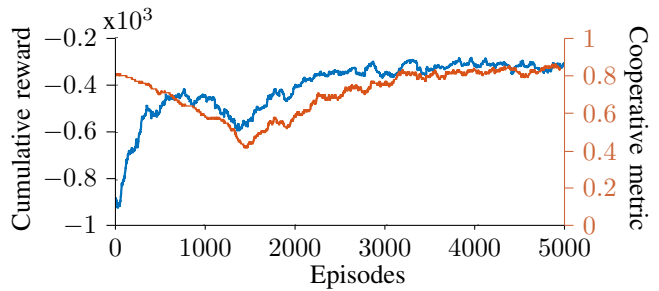
\begin{figure}[tb]
    \centering
        \begin{tikzpicture}[remember picture]
        \node[anchor=south west,inner sep=0] (image) at (0,0) {\includegraphics[width=0.335\textwidth, height=0.16\textwidth, trim={4.5cm 3cm 3.6cm 0cm}, clip]{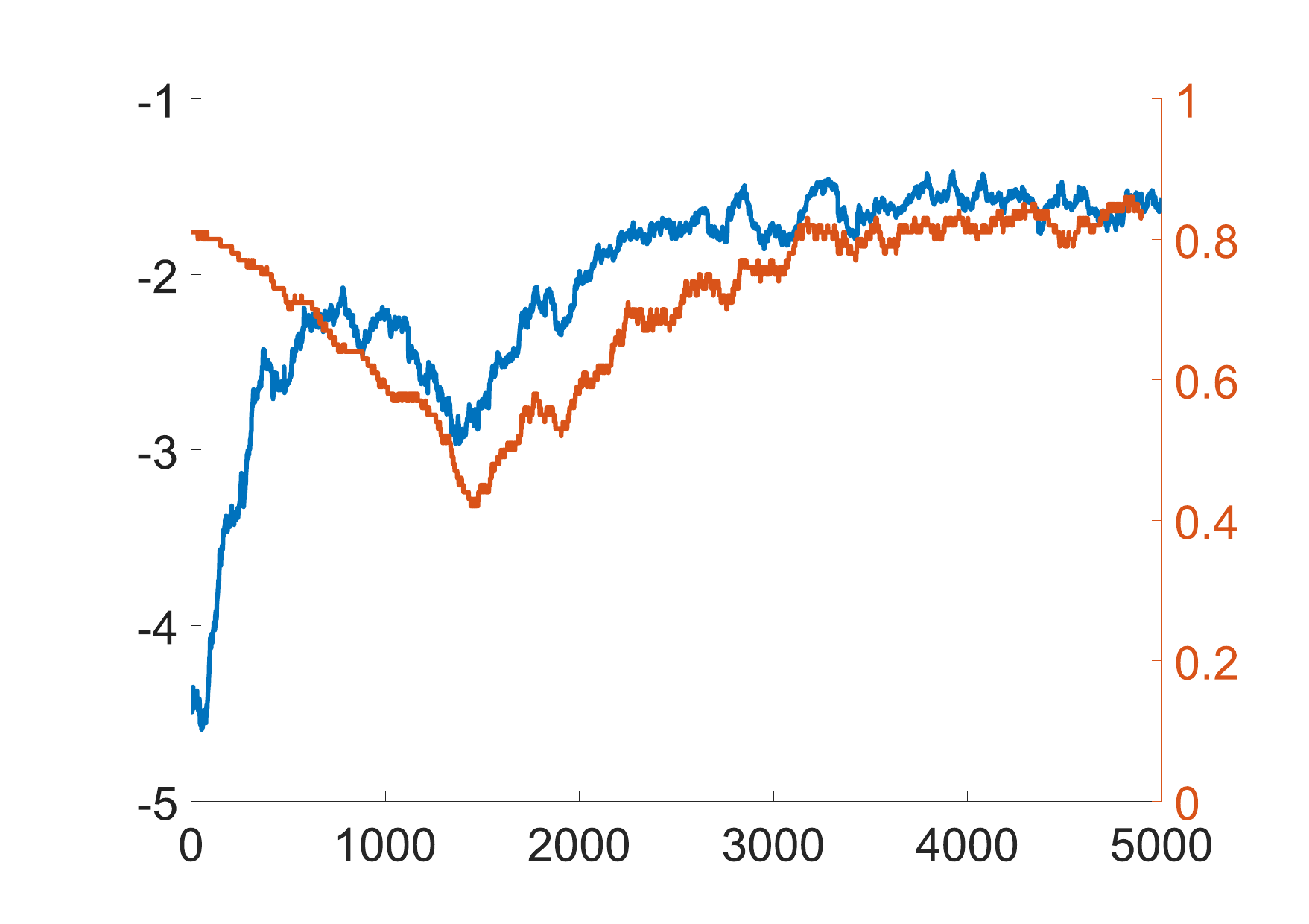}};
        \begin{scope}[x={(image.south east)},y={(image.north west)}]
            \draw[-] (0,0) -- (1,0) ;
            \draw[-] (0,0) -- (0,0.87) node[above] {\quad x$10^3$};
            \draw[-, color=myorange] (1,0.87) -- (1,0);

            \foreach \x/\xlabel in {0/0, 0.2/1000, 0.4/2000, 0.6/3000, 0.8/4000, 1/5000}
                \draw (\x,0) -- (\x,-0.02) node[below] {$\xlabel$};
            \foreach \y/\ylabel in {0/-1, 0.2175/-0.8, 0.435/-0.6, 0.6525/-0.4, 0.87/-0.2}
                \draw (0,\y) -- (-0.02,\y) node[left] {$\ylabel$};
            \foreach \y/\yylabel in {0/0, 0.174/0.2, 0.344/0.4, 0.522/0.6, 0.696/0.8, 0.87/1}
                \draw[myorange] (1,\y) -- (1.02,\y) node[right] {$\yylabel$};

            \node at (0.54,-0.24) {Episodes};
            \node[rotate=90]  at (-0.21,0.45) {Cumulative reward};
            \node[rotate=-90]  at (1.16,0.45) {Cooperative metric};
        \end{scope}
    \end{tikzpicture}
    \caption{Cumulative reward curve and Cooperative metric $CM$ of the training of the $n=1, \, m=5$ scenario. For readability purposes, the curve is smoothed with a moving average of $100$ steps (taken on the left) and we show only the first $5000$ episodes of the training.}
    \label{fig:nhmt_training}
\end{figure}

To further evaluate performance, we introduce the cooperative metric ($CM$) defined as the proportion of steps in an episode during which herders choose to pursue different targets. 
Figure \ref{fig:nhmt_training} shows both cumulative reward and $CM$ converging to steady values during training.
We note that, as training begins, herders select targets randomly; however, by applying a slow decay to the exploration rate $\epsilon$, they learn to complete the task but frequently choose the same targets, resulting in a low $CM$.
Furthermore, without any enforced coordination, the herders eventually learn to cooperate by avoiding the pursuit of the same targets, thereby maximizing the global reward. 
This improvement is reflected in the increase of both the cumulative reward and $CM$ in Figure \ref{fig:nhmt_training}, demonstrating effective learned cooperation.

To assess the impact of cooperation, we compare the policy trained in the $\hat{n}=2, \hat{m}=5$ case with two other scenarios: (i) herders independently selecting targets based on the policy learned in the $\hat{n}=1, \hat{m}=5$ scenario, and (ii) herders following the heuristic rule from \cite{auletta2022a}, which enforces cooperation by dynamically partitioning the plane into sectors, where each herder selects targets within its assigned sector (referred to as \textit{P2P} in \cite{auletta2022a}). 

\begin{figure}[tb]
    \centering
        \begin{tikzpicture}[remember picture]
        \node[anchor=south west,inner sep=0] (image) at (0,0) {\includegraphics[width=0.4\textwidth, height=0.16\textwidth, trim={4.5cm 3cm 0cm 0cm}, clip]{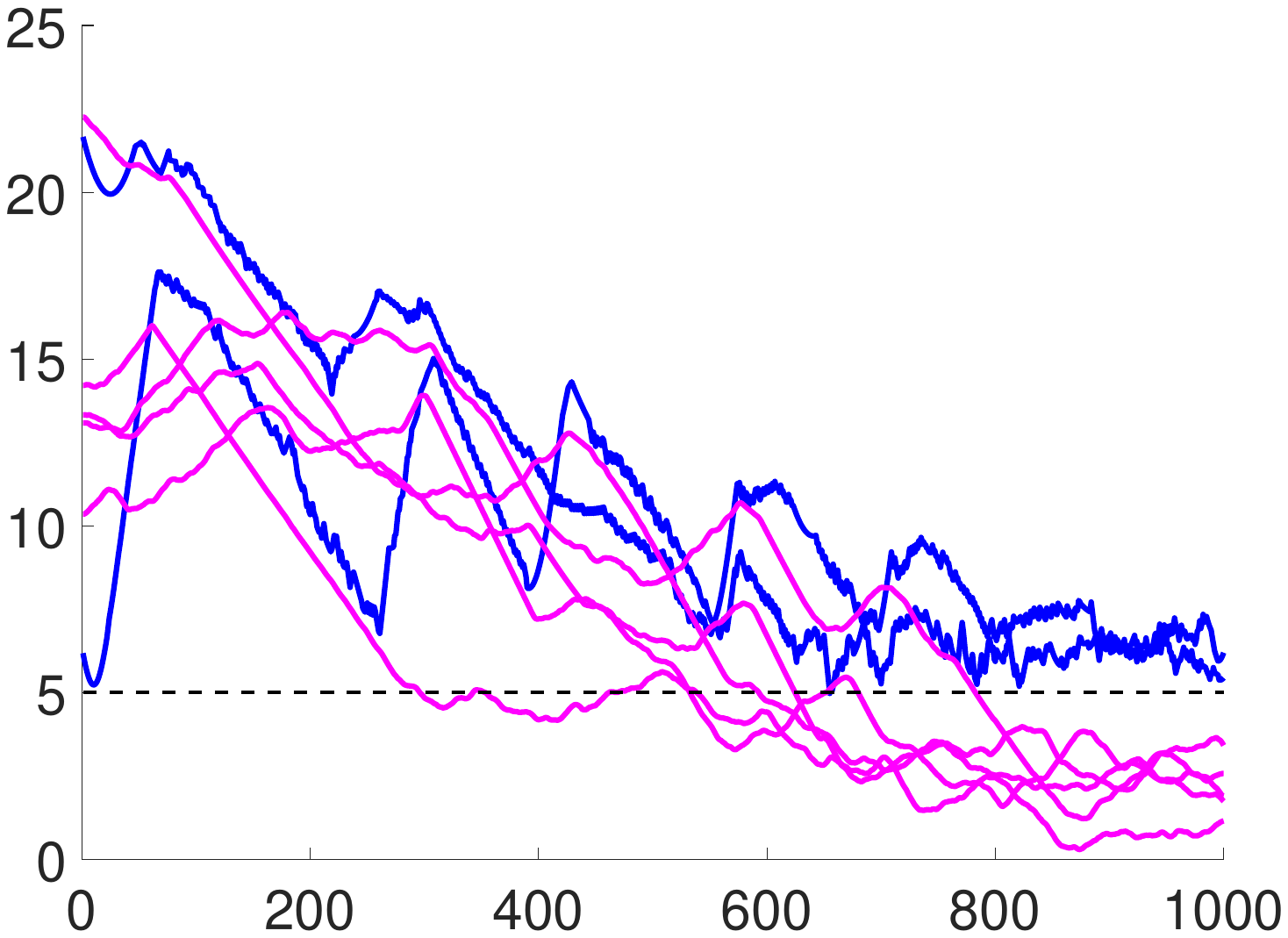}};
        \begin{scope}[x={(image.south east)},y={(image.north west)}]
            \draw[-] (0,0) -- (0.87,0) ;
            \draw[-] (0,0) -- (0,0.87) ;

            \foreach \x/\xlabel in {0/0, 0.174/200, 0.348/400, 0.522/600, 0.694/800, 0.87/1000}
                \draw (\x,0) -- (\x,-0.02) node[below] {$\xlabel$};
            \foreach \y/\ylabel in {0/0, 0.168/5, 0.348/10, 0.522/15, 0.694/20, 0.87/25}
                \draw (0,\y) -- (-0.02,\y) node[left] {$\ylabel$};

            \node at (0.45,-0.22) {Steps};
            \node[rotate=90]  at (-0.13,0.45) {Radii};
        \end{scope}
    \end{tikzpicture}
    \caption{Validation example of the proposed solution in the $n=2, m=5$ scenario. The radii of the herders (blue lines) and targets (magenta lines) are shown, compared to $\rho_G=5$ (black dotted line). 
    }
    \label{fig:nhmt_validation}
\end{figure}

Validation results in Table \ref{tab:validation},  show that our data-driven approach significantly outperforms the non-cooperative case and achieves performance comparable to state-of-the-art methods. 
For completeness, an example of the validation is shown in Figure \ref{fig:nhmt_validation}, where all targets enter and remain within the goal region, with $\norm{\mathbf{T}_i} \leq \rho_G$ for $i=1, \dots, 5$.

\begin{table}[tb]
    \centering
    \begin{tabular}{|c||c|c|c|}
        \hline
        Policy & Success \% & $\tau^{\star}$ (steps) & $CM$ \\
        \hline
        \hline
        \multicolumn{4}{|l|}{\textbf{Scenario n=1,  m=1}} \\
        \hline
        $\hat{n}=1, \hat{m}=1$ & $100$ & $682 \pm 320$ & -- \\
        \hline
        \multicolumn{4}{|l|}{\textbf{Scenario n=1, m=5}} \\
        \hline
        $\hat{n}=1, \hat{m}=5$ & $92.8$ & $10088 \pm 3961$ & -- \\
        \hline
        \multicolumn{4}{|l|}{\textbf{Scenario n=2, m=5}} \\
        \hline
        $\hat{n}=1, \hat{m}=5$ & $89.8$ & $10287 \pm 4044$ & $0.03$ \\
        $\hat{n}=2, \hat{m}=5$ & $100$ & $3197 \pm 1292$ & $0.94$ \\
        P2P \cite{auletta2022a} & $100$ & $2559 \pm 941$ & $1.00$\\
        \hline
    \end{tabular}
    \caption{Validation results (success rate, average settling time $\tau^\star$, and average Cooperative Metric $CM$) of the proposed solutions in the different scenarios. We compare in the full setting ($n=2, m=5$) the proposed $\hat{n}=2, \hat{m}=5$ policy to the case in which the two herders select the target independently using the policy trained in the one herder and five targets scenario ($\hat{n}=1, \hat{m}=5$ policy) and P2P, a rule-based target selection strategy proposed in \cite{auletta2022a}.}
    \label{tab:validation}
\end{table}

\paragraph{Robustness}
We varied the model parameters (i) $V_\R{max,T}$, (ii) $\sigma$, (iii) $c$, and (iv) $\lambda$ from the nominal values listed in Table \ref{tab:model_param}. Specifically, in each episode, each parameter was sampled from a Gaussian distribution centered on its nominal value $\mu$, with a standard deviation of $0.1\mu$. Using the same 1000 initial conditions, our solution achieves $99.8\%$ success rate, settling time of $\tau^{\star} = 3278 \pm 1338$ steps and $CM = 0.94$.  In such conditions, our solution achieves performances similar to the nominal scenario (see Table \ref{tab:validation}), showing robustness of the solution to parametric variations.

\subsection{Towards a scalable solution} \label{sec:scalability}
Since the number of agents used in training ($\hat{n}$ herders and $\hat{m}$ targets) dictates the structure of the neural network, our approach does not scale easily to different number of herders and targets. To tackle this limitation, 
we leverage the existing network trained with $\hat{n}$ herders and $\hat{m}$ targets. However, each herder is assumed to sense only up to the nearest $\hat{m}$ targets and  $\hat{n}-1$ herders, regardless of the actual number of agents.

\begin{figure}[tb]
    \centering
    \begin{minipage}{\textwidth}
        \begin{subfigure}[b]{0.245\textwidth}
            \centering
            \includegraphics[width=0.8\textwidth, trim={1.5cm 1cm 1.5cm 1cm}, clip]{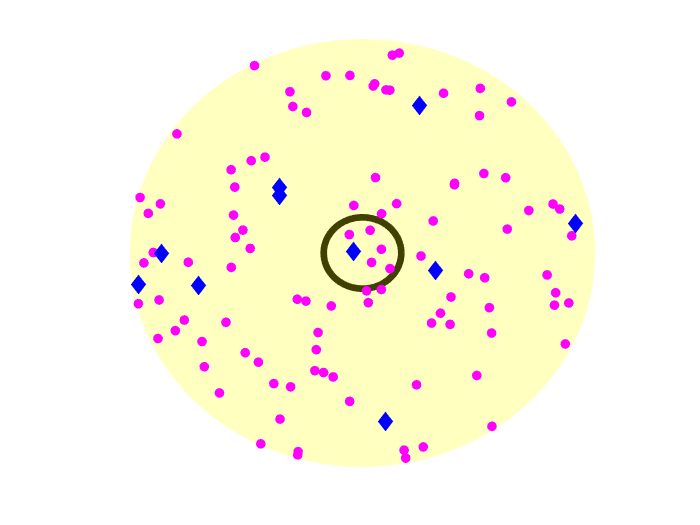}
            \caption{}
            \label{fig:validation_10h100t_IC}
        \end{subfigure}%
        \hspace{0.01cm}%
        \begin{subfigure}[b]{0.245\textwidth}
            \centering
            \includegraphics[width=0.8\textwidth, trim={1.5cm 1cm 1.5cm 1cm}, clip]{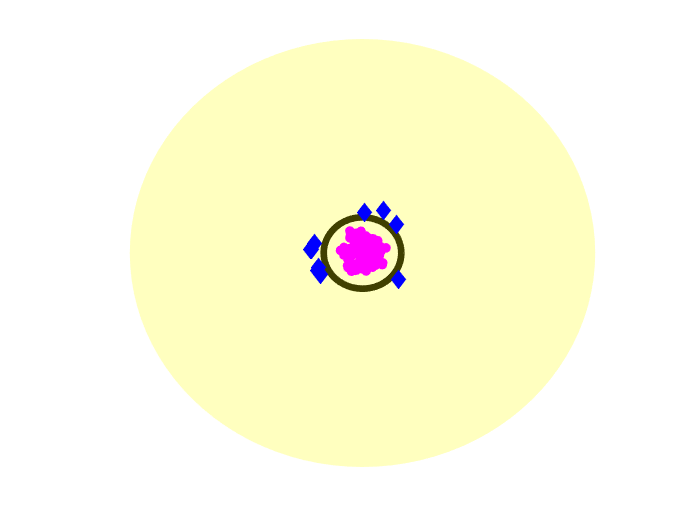}
            \caption{}
            \label{fig:validation_10h100t_FC}
        \end{subfigure}%
    \end{minipage}
    
    \vspace{0.1cm}

    \begin{subfigure}[t]{0.4\textwidth}
        \hspace{-0.6cm}
        \begin{tikzpicture}
        \node[anchor=south west,inner sep=0] (image) at (0,0) {\includegraphics[width=\textwidth, height=0.36\textwidth, trim={4.5cm 3cm 0cm 0cm}, clip]{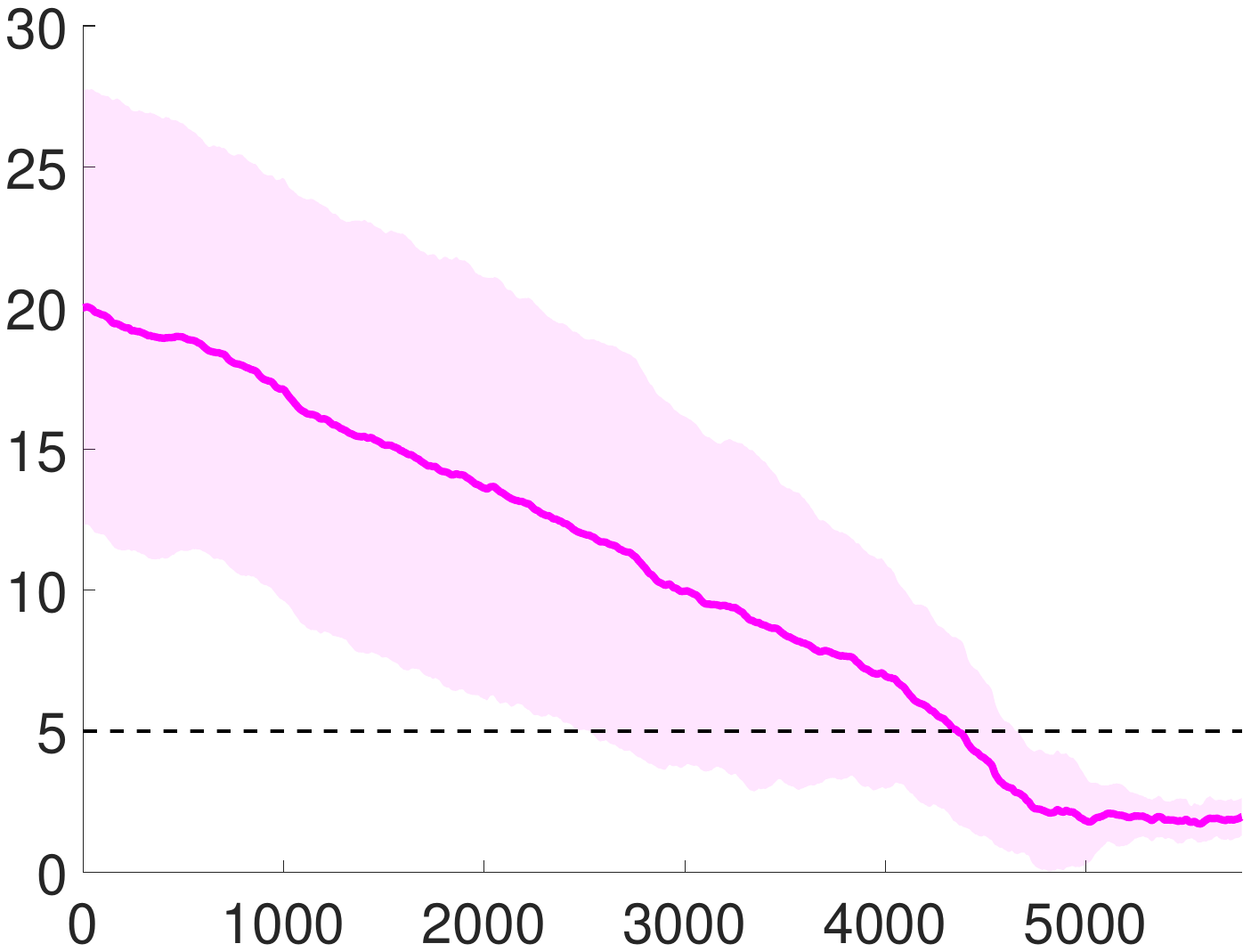}};
        \begin{scope}[x={(image.south east)},y={(image.north west)}]
            \draw[-] (0,0) -- (0.87,0) ;
            \draw[-] (0,0) -- (0,0.87) ;

            \foreach \x/\xlabel in {0/0, 0.15/1000, 0.3/2000, 0.45/3000, 0.6/4000, 0.75/5000}
                \draw (\x,0) -- (\x,-0.02) node[below] {$\xlabel$};
            \foreach \y/\ylabel in {0/0, 0.141/5, 0.29/10, 0.435/15, 0.58/20, 0.725/25, 0.87/30}
                \draw (0,\y) -- (-0.02,\y) node[left] {$\ylabel$};

            \node at (0.4,-0.24) {Steps};
            \node[rotate=90]  at (-0.13,0.45) {Targets' radii};
            \node[] at (0,1.1) {(a)};
            \draw[->] (0,1) -- ++(0, -0.1) node[below] {};
            \node[] at (0.87,0.3) {(b)};
            \draw[->] (0.87,0.2) -- ++(0, -0.1) node[below] {};

        \end{scope}
        \end{tikzpicture}
        \caption{}
    \end{subfigure}

    \caption{Validation example in the $n = 10$ herders and $m = 100$ targets scenario. Given the domain $\mathbb{R}^2$, we show (a) initial and (b) final positions of the herders (blue diamonds) and targets (magenta dots), and the goal region (black circle). The initial conditions are generated in the yellow domain. Panel (c) shows the evolution of the targets' radii mean (magenta line) and standard deviation (magenta shade) ant the goal region radius $\rho_\R{G}$ (black dotted line).}
    \label{fig:validation_10h100t}
\end{figure}

In Figure \ref{fig:validation_10h100t} we report an example in an extended spatial domain $\mathbb{R}^2$. In this settings, our DQN-based strategy (previously trained considering $\hat{n} = 2$ and $\hat{m} = 5$), manages to control $n = 10$ herders to successfully steer and contain $m = 100$ targets \cite{lama2024}. 
Future work will focus on developing herdability charts for this strategy, as in \cite{lama2024}.

\section{Conclusion}
We proposed a decentralized, learning-based solution to the shepherding control problem, focusing on non-cohesive targets and multiple learning herders. Our approach utilizes a two-layer control architecture based on (Multi-Agent) Deep Q-learning. Initially, we trained a single herder to guide and contain a target within the goal region in a single-agent scenario, and then extended this to multiple agents. Each herder independently selects and steers targets without communication, naturally learning to cooperate and complete the group task more efficiently—achieving faster completion without relying on heuristic model-based rules.

We scaled the solution from  environments with relatively fewer agents to larger scenarios by assuming a fixed number of sensed agents per herder, demonstrating scalability when targets do not vastly outnumber herders. We see this work as a starting point for approximating optimal control strategies for shepherding, aiming to tackle larger systems and more complex dynamics -- such as time-varying behaviors and heterogeneities -- by leveraging the benefits of reinforcement learning over model-based heuristic methods.

Future work will assess this method's effectiveness by varying the number of agents and incorporating limited sensing, such as a restricted field of view. Other potential extensions include considering obstacles in the environment and exploring higher-dimensional spaces.

\appendix
Here we report the set of parameters used for numerical experiments presented in this paper. Movies available at \href{https://tinyurl.com/258w4f7f}{https://tinyurl.com/258w4f7f}.
\begin{table}[htb]
    \centering
    \begin{tabular}{|c|c||c|c|}
        \hline
        \textbf{Parameter} & \textbf{Value} & \textbf{Parameter} & \textbf{Value} \\
        \hline
        $\Delta t$ & $0.05$ & $\sigma$ & $1$ \\
        $L$ & $100$ & $\zeta$ & $1$\\
        $V_\R{max,T}$ & $1$ & $\beta$ & $5$ \\
        $V_\R{max,H}$ & $5$ & $c$ & $20$ \\
        $\rho_\R{G}$ & $5$ & $\lambda$ & $2.5$ \\
        \hline
    \end{tabular}
    \caption{Parameters of the mathematical model simulated.}
    \label{tab:model_param}
\end{table}

\begin{table}[htb]
    \centering
    \begin{tabular}{|c||c||c|}
        \hline
        \textbf{Hyperparameter} & \textbf{$n=1, m=1$} & \textbf{$n=1, m=5$}\\
        \hline
        Learning rate $\alpha$ & 1e-4 & 1e-4\\
        Batch size & 64 & 32\\
        Discount rate $\gamma$ & 0.99 & 0.99\\
        Exploration rate $\epsilon$ & 0.1 & 0.05\\
        Exploration rate decay & -- & 1e-5 \\
        Maximum buffer size & 10000 & 50000\\
        Minimum buffer size & 1000 & 1000\\
        Target network update $C$ & 1 & 10\\
        Timescales ratio $n_\R{w}$ & -- & 100 \\
        \hline
        $k_\R{1}$ & 0.5 & -- \\
        $k_\R{2}$ & 1 & --\\
        $k_\R{3}$ & 20 & --\\
        $k_\R{4}$ & 50 & --\\
        $k_\R{5}$ & -- & 1 \\
        \hline
    \end{tabular}
    \caption{Hyperparameters and reward weights for the scenarios $n=1, m=1$ and  $n=1, m=5$.}
    \label{tab:hyperparameters}
\end{table}

\bibliographystyle{ieeetr}
\bibliography{refs.bib} 

\end{document}